\documentstyle[prd,preprint,aps,epsfig,floats]{revtex}
\draft
\newcommand{\beq}{\begin{equation}}
\newcommand{\eeq}{\end{equation}}
\newcommand{\bea}{\begin{eqnarray}}
\newcommand{\eea}{\end{eqnarray}}
\begin{document}

\title{Methods in the LO Evolution of Nondiagonal Parton 
Distributions: The DGLAP Case} 

\author{Andreas Freund, Vadim Guzey}

\address{
Department of Physics,  The Pennsylvania State University\\
University Park, PA  16802, U.S.A.}

\maketitle

\begin{abstract}
In this paper, we discuss the algorithms used in the LO evolution program
for nondiagonal parton distributions in the DGLAP region and discuss the 
stability of the code.
Furthermore, we demonstrate that we can reproduce the case of the LO diagonal 
evolution within  $0.5\%$ of the original code as developed by the 
CTEQ-collaboration.   
\end{abstract}

\section{Introduction}
\label{intro}

Due to the recent availability of exclusive hard diffraction data at HERA, 
there has been a great interest in the study of generalized parton 
distributions also known as nondiagonal, off-forward or non-forward parton 
distributions occurring in these reactions (see 
Ref.\ \cite{1,2,3,4,ours,6,7,8,9,MR97,mpw}). These parton distributions are 
different from the usual, diagonal distributions found in e.g.\ inclusive DIS 
since one has a finite
momentum transfer to the proton due to the exclusive nature of the reactions.
In this paper we give an exposition of the algorithms used to numerically solve
the generalized GLAP-evolution equations. The main part of the evolution
program was taken over from the CTEQ package for the diagonal parton 
distributions
from inclusive reactions. At this point in time the evolution kernels for
generalized parton distributions are known only to leading order in 
$\alpha_s$ and thus our analysis will be a leading order one.

The paper is organized in the following way. In Sec.\ \ref{def} we will quickly
review the formal expressions for the parton distributions and the evolution 
equations together with the explicit expressions for the kernels and a first 
comment on the arising numerical problems. In Sec.\ \ref{num1} we will explain
the difference of our algorithms to the ones used in the original CTEQ package
and then give a detailed account of how we implemented our algorithms. In 
Sec.\ \ref{num2} we demonstrate the stability of our code and show that we 
reproduce the case of the usual or diagonal parton distributions within $1\%$
for a vanishing asymmetry factor. Sec.\ \ref{concl} contains concluding
remarks.

\section{Review of Nondiagonal Parton Distributions, Evolution Equations and
Kernels}
\label{def}

\subsection{Nondiagonal Parton Distributions}
\label{ndpd}

\begin{figure}
\centering
\mbox{\epsfig{file=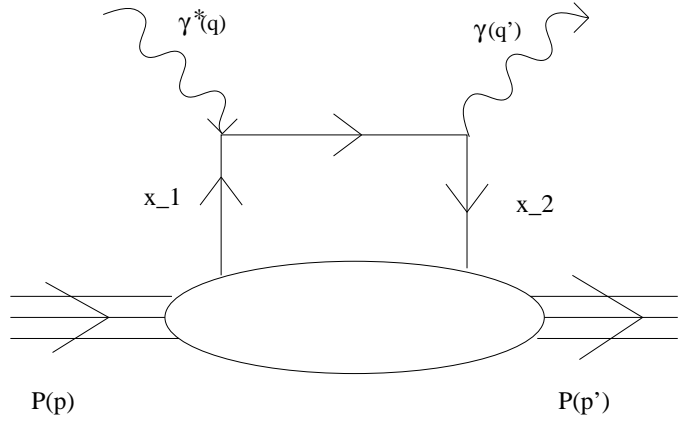,height=5.5cm}}
\caption{The lowest order handbag contribution to DVCS with $Q^2=-q^2$ and 
$q'^2=0$.}
\label{hand}
\end{figure}

Generalized or, from now on nondiagonal parton distributions, occur for example
in exclusive, hard diffractive $J/\psi$ or $\rho$ meson production and 
alternatively in deeply virtual Compton scattering (DVCS), where a real photon
is produced. As mentioned in Sec.\ \ref{intro} since one imposes the condition
of exclusiveness on top of the diffraction condition, one has a kinematic 
situation in which there is a non-zero momentum transfer onto the target 
proton as evidenced by the lowest order ``handbag'' diagram of DVCS in 
Fig.\ \ref{hand}.
The picture serves to only introduce the kinematic notations used throughout
the text and nothing more. For more on DVCS see for example 
Ref.\ \cite{6,7,muell,Ji1,BM,DGPR,chen,fas,man,ca,jios}.

The nondiagonal quark and gluon distributions have the following formal
definition as matrix elements of bilocal, path-ordered and renormalized quark 
and gluon operators sandwiched between different momentum states of the proton
as in the factorization theorems for exclusive vector meson 
production \cite{4} and DVCS \cite{7,ca,jios}:
\bea
& &f_{q/p}= \int^{\infty}_{-\infty}\frac{dy^-}{4\pi}e^{-ix_2p^+y^-}\langle p|
T\bar \psi(0,y^-,{\bf 0_{\perp}})\gamma^+{\it P}\psi(0)|p'\rangle,\nonumber\\
& &f_{g/p}= -\int^{\infty}_{-\infty}\frac{dy^-}{2\pi}\frac{1}{x_1x_2p^+}
e^{-ix_2p^+y^-}\langle p|
T G_{\nu}^+(0,y^-,{\bf 0_{\perp}}){\it P}G^{\nu+}(0)|p'\rangle.
\eea
with $x_2 = x_1 - \Delta$ where the asymmetry or nondiagonality parameter 
$\Delta$ is usually $x_{Bj}$ in, for example, DVCS or exclusive vector meson
production however not in diffractive di-muon production.
   
\subsection{The GLAP-Evolution Equations for Nondiagonal Parton Distributions}
\label{ee}

The GLAP-evolution equations follow from the usual renormalization group 
invariance of the factorization formula and lead to the following evolution
equations for the singlet(S) and non-singlet(NS) case \cite{ours,6,7}:
\bea
& &\frac{dq_{NS}(x_{1},\Delta,Q^2)}{d\ln{Q^2}}=\int^{1}_{x_{1}} 
\frac{dy_{1}}{y_{1}}P_{qq}q_{NS}(y_{1},\Delta,Q_0^2),\nonumber\\
& &\frac{dg_S(x_{1},\Delta,Q^2)}{d\ln{Q^2}}=\int^{1}_{x_{1}} 
\frac{dy_{1}}{y_{1}}\left [ P_{gg}g_S(y_{1},\Delta,Q_0^2)+P_{gq}q_S(y_{1},
\Delta,Q_0^2)\right ],\nonumber\\
& &\frac{dq_S(x_{1},\Delta,Q^2)}{d\ln{Q^2}}=\int^{1}_{x_{1}} 
\frac{dy_{1}}{y_{1}}\left [ P_{qq}q_S(y_{1},\Delta,Q_0^2)+P_{qg}g_S(y_{1},
\Delta,Q_0^2)\right ].
\label{eq:evol}
\eea
Note that $q_{S,NS} = x_1 f_{q/p}$, $g_S = x_1x_2f_{g/p}$ and the kernels 
to leading order\cite{f1} are given by \cite{ours,6,7}:
\bea
P_{qq,S,NS}(x_1,\Delta)&=&\frac{\alpha_s}{\pi} C_f\left [ \frac{x_1 + 
x_1^{3} - \Delta (x_1 + x_1^{2})}
{(1 - \Delta) (1 - x_1)} -\delta (1-x_1)\left [\int^{1}_{0} 
\frac{dz_1}{z_1} + \int^{1}_{0}\frac{dz_2}{z_2} -\frac{3}{2} 
\right ] \right ],\nonumber\\
P_{qg,S} (x_1, \Delta) &=& \frac{\alpha_s}{\pi} N_F \frac{[x_1^{3} + 
x_1(1 - x_1)^{2} - x_1^{2} \Delta]}{(1 - \Delta)^2},\nonumber\\
P_{gq,S}(x_1, \Delta) &=& \frac{\alpha_s}{\pi} C_F \frac{[ 1 + (1-x_1)^{2} - 
\Delta]}{1 - \Delta},\nonumber\\
P_{gg,S}(x_1, \Delta)&=&N_{c} [2\frac{(1 - x_1)^2 + (\frac{1}{2} - x_1^{2})
(x_1 - \Delta)}{(1 - \Delta)^2} - 1 - x_1 + 
\frac{1}{1 - x_1} + \frac{x_1 - \Delta}
{(1 - x_1)(1 - \Delta)}\nonumber\\  
& & + \delta (1-x_1) \left [ \frac{\beta_0}{2N_C} - \int^{1}_{0} 
\frac{dz_1}{z_1} - \int^{1}_{0} \frac{dz_2}{z_2} \right ]].
\label{eq.kern}
\eea
With our definitions, we obtain for the diagonal limit, i.e.\ , $\Delta=0$,
$q_{S,NS}\rightarrow xQ(x,Q^2)$ and $g_S\rightarrow xG(x,Q^2)$ where $Q$ and 
$G$ are the usual parton densities.

A word concerning the above employed regularization prescription which is the 
usual
+ - prescription in the first integral below and a generalized 
+ - prescription for the second integral, is in order, since these
prescriptions have direct implications on the numerical treatment of the 
integrals involved. In convoluting the 
above kernels, after appropriate scaling of $x_1$ and $\Delta$ with $y_1$,
with a nondiagonal parton density, one has to replace $z_1$ and $z_2$ in the 
regularization integrals with $z_1 \rightarrow (y_1 - x_1)/y_1$ and $z_2 
\rightarrow (y_1-x_1)/(y_1-\Delta)$. This leads to the following 
regularization prescription as employed in our modified version
of the CTEQ package and in agreement with Ref.\ \cite{7}:
\bea
\int^{1}_{x_1}\frac{dy_1}{y_1}\frac{f(y_1)}{1-x_1/y_1}_{+} &=& 
\int^{1}_{x_1}\frac{dy_1}{y_1}\frac{y_1f(y_1)-x_1f(x_1)}{y_1-x_1} 
+ f(x)\ln (1-x_1)\\
\int^{1}_{x_1}dy_1\frac{(x_1-\Delta)f(y_1)}{(y_1-x_1)(y_1-\Delta)}_{+} &=& 
\int^{1}_{x_1}\frac{dy_1}{y_1}\frac{y_1f(y_1)-x_1f(x_1)}{y_1-x_1}  - 
\int^{1}_{x_1}\frac{dy_1}{y_1}\frac{y_1f(y_1)-\Delta f(x_1)}
{y_1-\Delta}\nonumber\\
& & + f(x_1)\ln \left ( \frac{1-x_1}{1 - \Delta} \right )
\label{reg}
\eea

Eq.\ \ref{reg} and a closer inspection of Eq.\ \ref{eq.kern} reveals that 
if one were to integrate each term by itself one would encounter 
infinities in all the 
expressions at both the lower bound of integration if $\Delta = y_1$ and in
taking the limit $\Delta = x_1$. Although Eq.\ \ref{eq.kern} is completely 
analytical, it will cause numerical problems since the cancellations of the 
infinite terms can only be done in the analytical expressions. This is in 
contrast to the diagonal case where such problems are absent. The integration 
over $Q^2$ is identical to the diagonal case and hence has already been dealt
with in the original CTEQ-code.  

\section{Differences between the CTEQ and our Algorithms}
\label{num1}

Let us point out in the beginning that our code is to $99\%$ the original
CTEQ-code (for a detailed account of this code see Ref.\ \cite{cteq}). 
We only modified the subroutines NSRHSM, NSRHSP and SNRHS within
the subroutine EVOLVE and added the subroutines NEWARRAY and NINTEGR. These
routines are only dealing with the convolution integrals but not with, for 
example, the $Q^2$-integration or any other part of the CTEQ-code which remains
unchanged. This is due to the fact that the main difference between the 
diagonal and nondiagonal evolution stems from the different kernels which only
influence the convolution integration and nothing else. 

In order to make the simple changes in the existing routines more obvious we 
will first deal with the new subroutines.

\subsection{NEWARRAY and NINTEGR}
\label{nsubs}

Due to the increased complexity of the convolution integrals as compared to
the diagonal case as pointed out in Sec.\ \ref{ee}, we were forced to 
slightly change the very elegant and fast integration routines employed
in the original CTEQ-code. The basic idea, very close to the one in the 
CTEQ-code, is the following:
Within the CTEQ package, the parton distributions are given on a dynamical 
$x$- and 
$Q$-grid of variable size where the convolution of the kernels with the initial
distribution is performed on the $x$-grid. Due to the possibility of singular 
behavior of the integrands, we perform the convolution integrals by first 
splitting up the region of integration according to the number of grid 
points in $x$, analytically integrating between two grid points $x_i$ and 
$x_{i+1}$ where $i$ runs from $1$ to the specified number of points in $x$ 
and then adding up the contributions from the small intervals as exemplified 
in the following equation:
\beq
\int^1_{x_1}\frac{dy_1}{y_1}f(x_1/y_1,\Delta/y_1,y_1) = \Sigma_{i=0}^{N}
\int^{x_{i+1}}_{x_i}\frac{dy_1}{y_1} f(x_1/y_1,\Delta/y_1,y_1),
\label{split}
\eeq
where $f(x_1/y_1,\Delta/y_1,y_1)$ is the product of the initial distribution 
for each evolution
step and an evolution kernel with $x_0=x_1$, $x_N=1$. We can do the 
integration analytically between two neighboring grid points by approximating 
the distribution function $f(y_1)$ through a second order polynomial 
$ay_1^2 + by_1 +c$, using the fact that we know the function on the grid points 
$x_{i-1},x_i$ and $x_{i+1}$ and can thus compute the coefficients a,b,c of 
the polynomial in the following way, given the function is well behaved and 
the neighboring grid points are close together \cite{f2}:
\bea
f(x_{1+1}) &=& ax^2_{i+1}+bx_{i+1}+c\nonumber\\
f(x_i) &=& ax^2_i+bx_i +c\nonumber\\
f(x_{i-1}) &=& ax^2_{i-1}+bx_{i-1}+c
\eea
which yields a $3\times 3$ matrix relating the coefficients of the polynomial 
to the values of the distribution functions at $x_{i-1},x_i$ and $x_{i+1}$.
Inverting this matrix in the usual way one obtains a matrix relating the $x$
values of the distribution function to the coefficients making it possible to 
compute them just from the knowledge of the different $x$ values and the 
value of the distribution function at those $x$ values. This calculation is 
implemented in NEWARRAY where the initial distribution is handed to the 
subroutine and the coefficient array is then returned. The coefficient array
in which the values of the coefficients for the integration
are stored, has $3$ times the size of the user-specified
number of points in $x$ since we have $3$ coefficients for each bin in $x$.  
We treat the last integration between the points $x_0$ and $x_1$ 
again by 
approximating the distribution in this last bin through a second order 
polynomial. However, for this last bin, the coefficients are computed using
the last three values in $x$ and of the distribution at those points, since the 
point $x_{-1}$ which would be required according to the above prescription for
calculating the coefficients, does not exist.

After having regrouped the terms appearing in the convolution integral in such
a way that all the necessary cancellations of large terms occur within the 
analytic expression for the integral and not between different parts of the 
convolution integral, the integration of the different terms is performed in 
the new subroutine NINTEGR with the aid of the coefficient array from NEWARRAY.
As mentioned above the convolution integral from $x_1$ to $1$ is split up 
into several intervals in which the integration is 
carried out analytically. To give an example of this procedure we consider 
the convolution integral of $P_{qg}(x_1/y_1,\Delta/y_1)$ with the 
parton distribution $g_S(y_1)$:
\beq
\int^1_{x_1}\frac{dy_1}{y_1}P_{qg}g_S = \int^1_{x_1}\frac{dy_1}{y_1}
\frac{x^2_1(x_1-\Delta)}{y_1(y_1-\Delta)^2}g_S(y_1)+\int^1_{x_1}
\frac{dy_1}{y_1}\frac{x_1(y_1-x_1)^2}{y_1(y_1-\Delta)^2}g_S(y_1) 
\label{examp}
\eeq
suppressing presently irrelevant factors in front of the integral. 
The two parts
in Eq.\ \ref{examp} are calculated in different parts of NINTEGR
and then put together in either NSRHSM, NSRHSP or SNRHS. In NINTEGR the 
integrals are split up according to Eq.\ \ref{split} and then 
analytically evaluated in the different $x$-bins \cite{f3}. 
If the dependence of
the integrand on $\Delta$ is only of a multiplicative nature it is enough
to compute the integral for each bin once. To get the value of 
convolution integral for a term with such a $\Delta$ \cite{f4} dependence, 
it is enough 
to store the result of the integration in the bin from $x_{N-1}$ to $x_N$
in the output array for this term at the position $N-1$ \cite{f5}, 
add to this result the value of the integral in the bin from $x_{N-2}$ to 
$x_{N-1}$ and store it at the 
position $N-2$ and so forth. In this manner one only has to calculate $N-1$
integrals, however if the integrand has a more complicated dependence on
$\Delta$ like $x_1-\Delta$ one needs to compute $N(N-1)/2$ integrals. 
For example in order to find the integration value for the $x_{N-1}$ bin
with $x_1=x_{N-1}$ one needs only one integral but at $x_{N-2}$ we have to 
redo our integral for the $x_{N-1}$ bin since $x_1=x_{N-2}$ plus we need to 
add the contribution from the $x_{N-2}$ bin to get the correct answer for the
output array at position $N-2$ and so forth. This need for additional
evaluations of integrals slows the program down but in the end it turns
out to be only about a factor of $4-5$ slower ( as tested on a SPARC 10 ) 
than the original CTEQ-code 
which is speed optimized. The integral with the regular + - prescription
is evaluated using the routine HINTEG from the original CTEQ-code whereas the 
generalized + - prescription is evaluated according to the methods described
above due to its nontrivial dependence on $x_1$ and $\Delta$. 

The case $x_1 = \Delta = x$ and $\Delta<< x_1$, are implemented in NINTEGR
in the same way as above but separately from each other and from
the more general case. For $x_1 = \Delta = x$ the form of the 
integrands simplify in such a way that one can use the integration routines 
INTEGR and HINTEG from the original CTEQ-code. In the case of $\Delta<<x_1$ 
the analytic expressions obtained for the above general case are expanded to 
first order in $\Delta$ and then the same methods as above for evaluating the 
integrals are applied. The last case also allows us to go to the diagonal case
by setting $\Delta=0$ without using the integration routines from the 
original CTEQ-code giving us a valuable tool to compare our code to the 
original one.

\subsection{Modifications in NSRHSM, NSRHSP and SNRHS}
\label{mod}

The modification in the already existing routines NSRHSM, NSRHSP and SNRHS 
of the original CTEQ package are 
rather trivial. The most notable difference is that the subroutine NEWARRAY 
is called every time either of the three subroutines is called since the 
distribution function handed down on an array changes with every call of
NSRHSM, NSRHSP and SNRHS. In NSRHSM and NSRHSP, NEWARRAY is only called once
since one is only dealing with the non-singlet part containing no gluons, 
whereas in SNRHS the subroutine for the singlet case, one needs a coefficient
array for both the quark and the gluon. Besides this change, the calls for 
INTEGR are replaced by NINTEGR according to how the convolution integral has 
been regrouped as explained in Sec.\ \ref{nsubs}. The different regrouped 
expressions are then added, after integration for different $x$-values, to obtain the 
final answer in an 
output array which is handed back to the subroutine EVOLVE. 
The method is the same as in the 
original CTEQ-code but the terms themselves have changed of course.

\section{Code Analysis}  
\label{num2}

As a first step we tested the stability and speed of convergence of the code 
and found that by doubling the number of points in the $x$-grid, which is only
relevant for the convolution integral, from $50$ to $300$ the result of our 
calculation changed by less than $0.5\%$, hence we can assume that our code 
converges rather rapidly. We also found the code to be stable down to an 
$x_2=10^{-10}$ beyond which we did not test. Furthermore we can reproduce
the result of the original CTEQ-code, i.e.\ the diagonal case in LO within 
$0.5\%$ giving us confidence that our code works well since the analytic
expressions for the diagonal case are the expansions of the general case 
of non vanishing asymmetry up to, but not including, $O(\Delta^2)$.  

In the following figures (Fig.\ \ref{nddratio}-\ref{nddratio5}) 
we compare, for illustrative purposes, the diagonal and nondiagonal case by 
plotting the ratio
\begin{eqnarray}
R_g(x_1,x_2,Q^2) = \frac{g(x_1,x_2,Q^2)}{x_1G(x_1,Q^2)}\nonumber\\
R_q(x_1,x_2,Q^2) = \frac{q(x_1,x_2,Q^2)}{x_1Q(x_1,Q^2)},
\end{eqnarray}
for various values of $x_1$, $Q^2$ and $\Delta=x_{Bj}$ \cite{f6}
, i.e.\ varying $x_2$, using the CTEQ4M and CTEQ4LQ \cite{f7}
 parameterizations \cite{cteq4}. We assume  the 
same initial conditions for the diagonal and nondiagonal case 
(see Ref.\ \cite{ours} for a detailed physical motivation of this ansatz).

\begin{figure}
\centering
\mbox{\epsfig{file=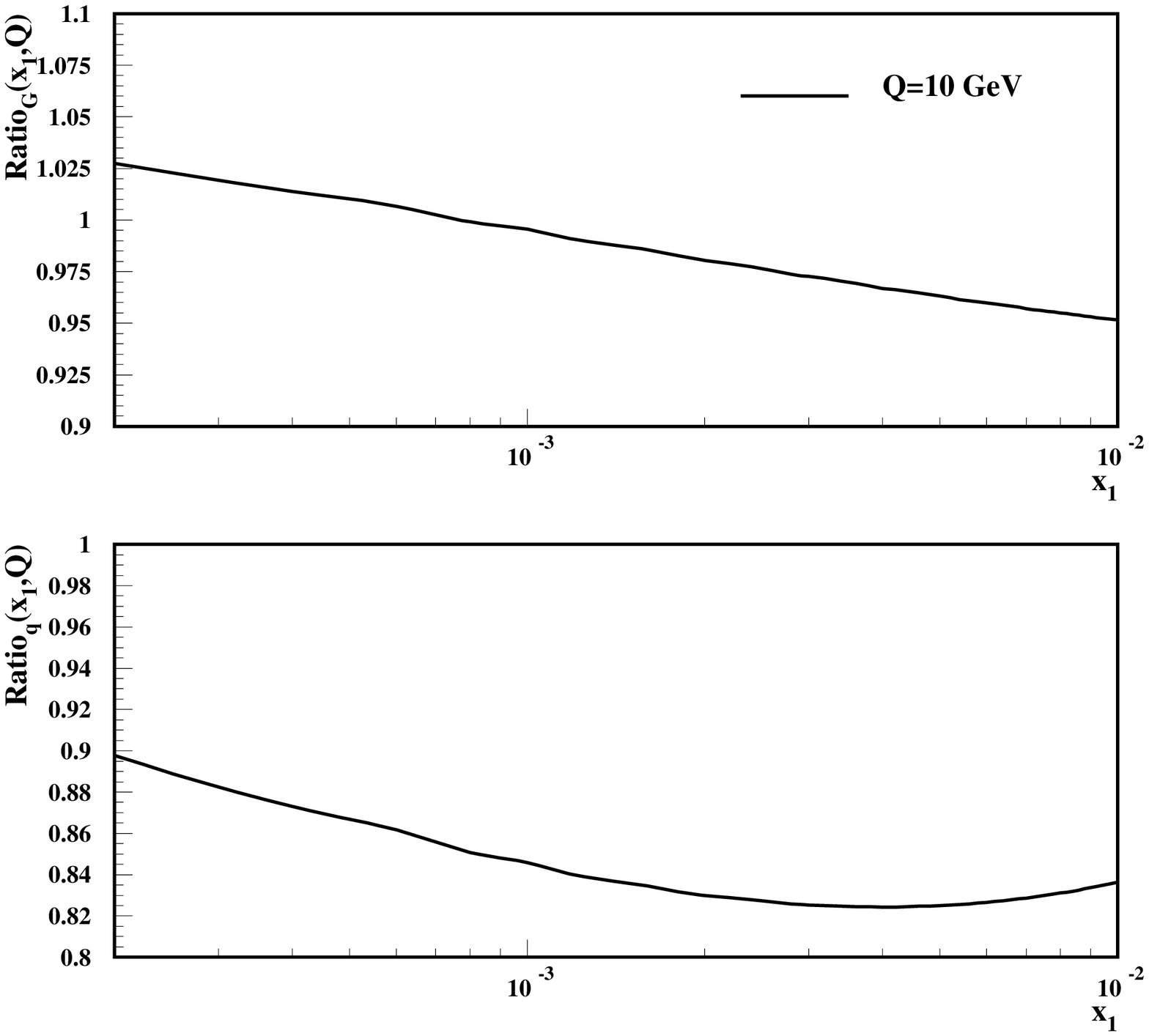,width=14cm,height=20cm}}
\vspace*{5mm}
\caption{The ratios for CTEQ4M to CTEQ4LQ for gluons and quarks 
in the diagonal case is plotted to demonstrate the difference between the 
LO evolution for these 
parameterizations.}
\label{nddratio6}
\end{figure}

The reader might wonder why only CTEQ4M and CTEQ4LQ and not GRV or MRS were 
used. The answer is not 
a prejudice of the authors against GRV or MRS but rather the fact that a 
comparison of CTEQ4M and CTEQ4LQ shows the same characteristic as comparing,
for example, CTEQ4M and GRV at LO. The observation is the following:
CTEQ4LQ is given at a different, rather low, $Q$, as compared to CTEQ4M and
hence one has significant corrections from NLO terms in the evolution at 
these scales. This leads to a large difference between CTEQ4LQ and CTEQ4M 
(see Fig.\ \ref{nddratio6}), if one evolves the CTEQ4LQ set 
from its very low $Q$ scale to the scale at which the CTEQ4M distribution is
given, making a sensitivity study of nondiagonal parton distributions for
different initial distributions impossible at LO. Of course, the inclusion
of the NLO terms corrects this difference in the diagonal case but since there
is no NLO calculation of the nondiagonal case available yet, a study of the
sensitivity of nondiagonal evolution to different initial distributions has to
wait.

The figures themselves suggest the following. The lower the starting 
scale, the stronger the effect of the difference of the nondiagonal evolution
as compared to the diagonal one and also that most of the difference between
nondiagonal and diagonal evolution stems from the first few steps in the 
evolution at lower scales. Secondly, under the assumption that the NLO 
evolution in the nondiagonal case will yield the same results for the parton
distributions at some scale $Q$, irrespective of the starting scale $Q_0$, in 
analogy to the diagonal case. One can say that the NLO corrections to the 
nondiagonal evolution will be in the same direction and same order of 
magnitude as the diagonal NLO evolution. If, in the nondiagonal case, the NLO 
corrections were in the opposite direction, which would lead to a 
marked deviation from the LO results, compared to the diagonal
case, the overall sign of the NLO nondiagonal kernels would have to change for
some $\Delta \neq 0$ since in the limit $\Delta \rightarrow 0$ we have to 
recover the diagonal case. This occurance is not likely for the following 
reason: First, the Feynaman diagrams involved in the calculation of the NLO 
nondiagonal kernels are the same as in the diagonal case, except for the 
different kinematics, therefore, we have a very good idea about the type 
of terms appearing in the kernels, namely polynomials, logs and terms in need 
of regularization such as $\ln(z) \frac{\ln (1-z)}{(1-z)}$. Moreover, the 
kernels, as 
stated before, have to reduce to the diagonal case in the limit of vanishing 
$\Delta$ which fixes the sign of most terms in the kernel, thus the only type
of terms which are allowed and could change the overall sign of the kernel are
of the form 
\beq
\frac{\Delta}{y_1} f(x_1/y_1,\Delta/y_1) 
\label{term}
\eeq
which will be numerically 
small unless $y_1 \simeq \Delta$ in the convolution integral of the 
evolution equations. Moreover, we know that in this limit the contribution of 
the regularized terms in the kernel give the largest contributions in the 
convolution integral and therefore sign changing contributions in the 
nondiagonal case would have to originate from regularized terms. This in turn
disallows a term like Eq.\ \ref{term} due to the fact that regularized terms
are not allowed to vanish in the diagonal limit, since the regularized 
 terms arise from the same Feynman diagrams in the both  
diagonal and nondiagonal case. Therefore, the overall sign of the contribution
of the NLO nondiagonal kernels will be the same as in the diagonal case.  

A word should be said about how the results of Ref.\ \cite{MR97} compare to 
ours. For the case of the same $\Delta =10^{-3}$ similar starting scales and 
almost identical values of $Q$ we find good agreement with their 
numbers for $R_g$ at $x_1\simeq \Delta$ \cite{f8}
and are slightly higher at larger $x_1$. The observed differences are due to 
the fact that the 
quark distributions are included in our evolution as compared to \cite{MR97}
and their initial distributions is slightly different.
We also find very similar ratios to \cite{MR97} if one changes the starting
scale to a lower one. The slight difference of a few percent in the ratios 
between us and \cite{MR97} can again be attributed to the fact that they used 
the GRV distribution as compared to our use of the CTEQ4 distributions, 
hence a slight difference in the starting scales and their lack of 
incorporating quarks into the evolution.

\section{Conclusions}
\label{concl}

We modified the original CTEQ-code in such a way that we can now compute
the evolution of nondiagonal parton distributions to LO. We gave a detailed
account of the modifications and the methods employed in the new or modified
subroutines. As the reader can see, the modifications and methods themselves 
are not something magical but rather a straightforward application of well
known numerical methods. We further demonstrated the rapid convergence and
stability of our code. In the limit of vanishing asymmetry we reproduce
the diagonal case in LO as obtained from the original CTEQ-code within 
$1\%$.  We also have good agreement with the results in Ref.\ \cite{MR97}.
In the future, after the NLO kernels for the nondiagonal case have
been calculated, we will extend the code to the NLO level to be on par with
the diagonal case. 

\section*{Acknowledgments}

This work was supported in part by the U.S.\ Department of Energy
under grant number DE-FG02-90ER-40577.
We would like to thank John Collins and Mark Strikman for helpful
conversations.

\begin{figure}
\centering
\vskip-3cm
\mbox{\epsfig{file=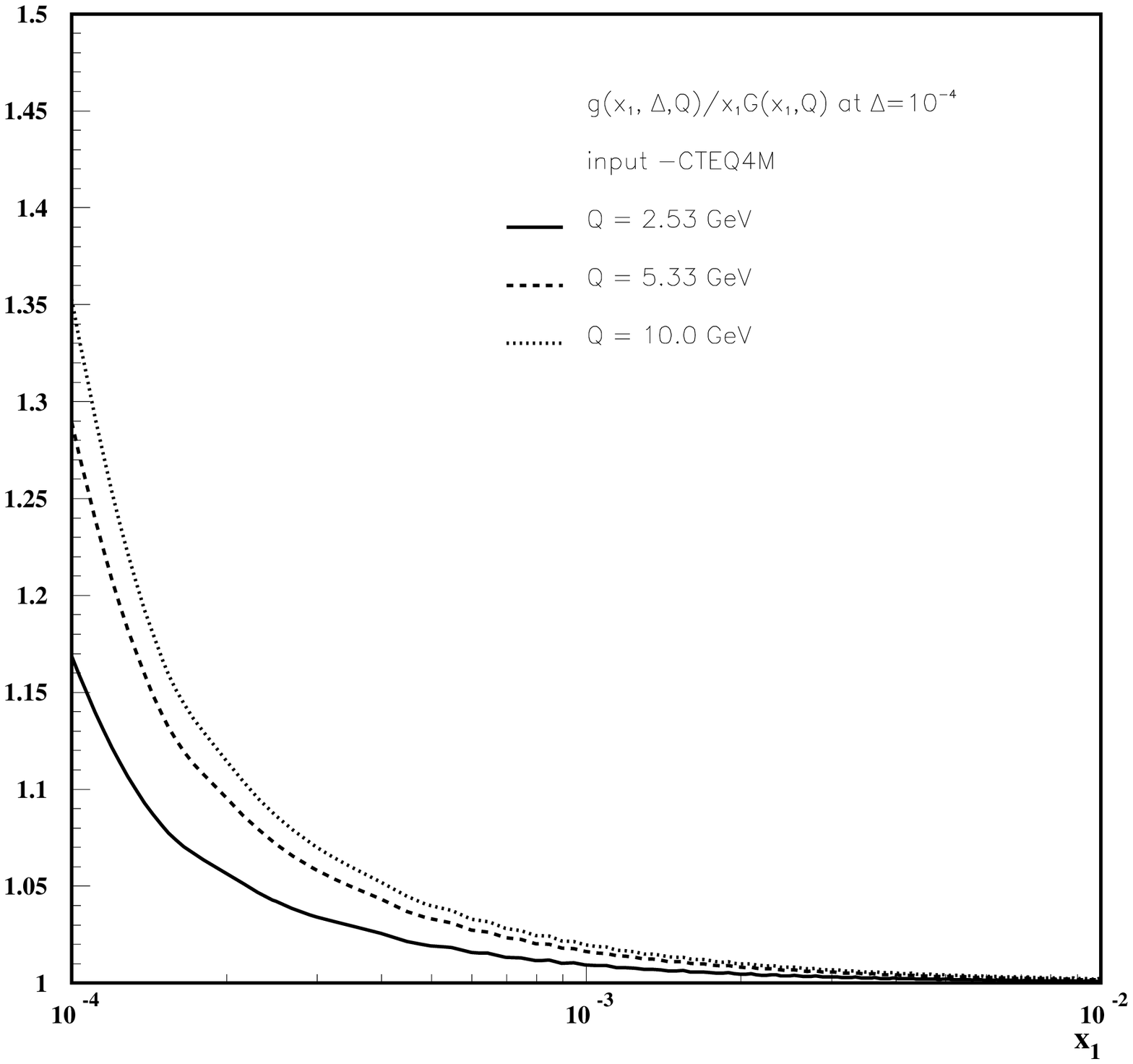,width=14cm,height=15cm}}
\vskip-3cm
\mbox{\epsfig{file=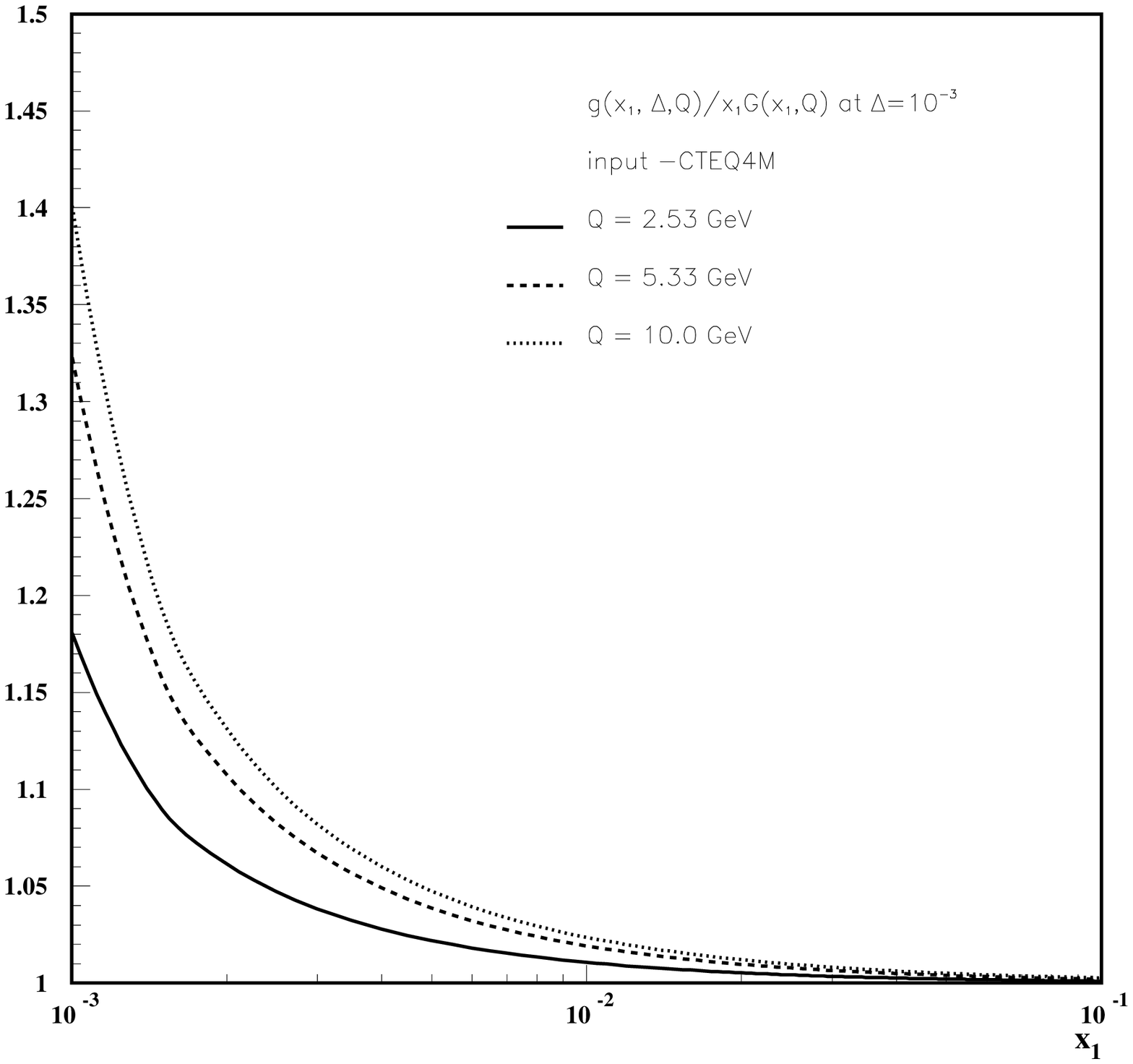,width=14cm,height=15cm}}
\vskip-2cm
\caption{$R_g$ is plotted versus $x_1$ for fixed $\Delta$ using the CTEQ4M 
parameterization with $Q_0=$\mbox{1.6 GeV} and $\Lambda$ =\mbox{ 202 MeV}.}
\label{nddratio}
\end{figure}
\newpage
\begin{figure}
\centering
\vskip-3cm
\mbox{\epsfig{file=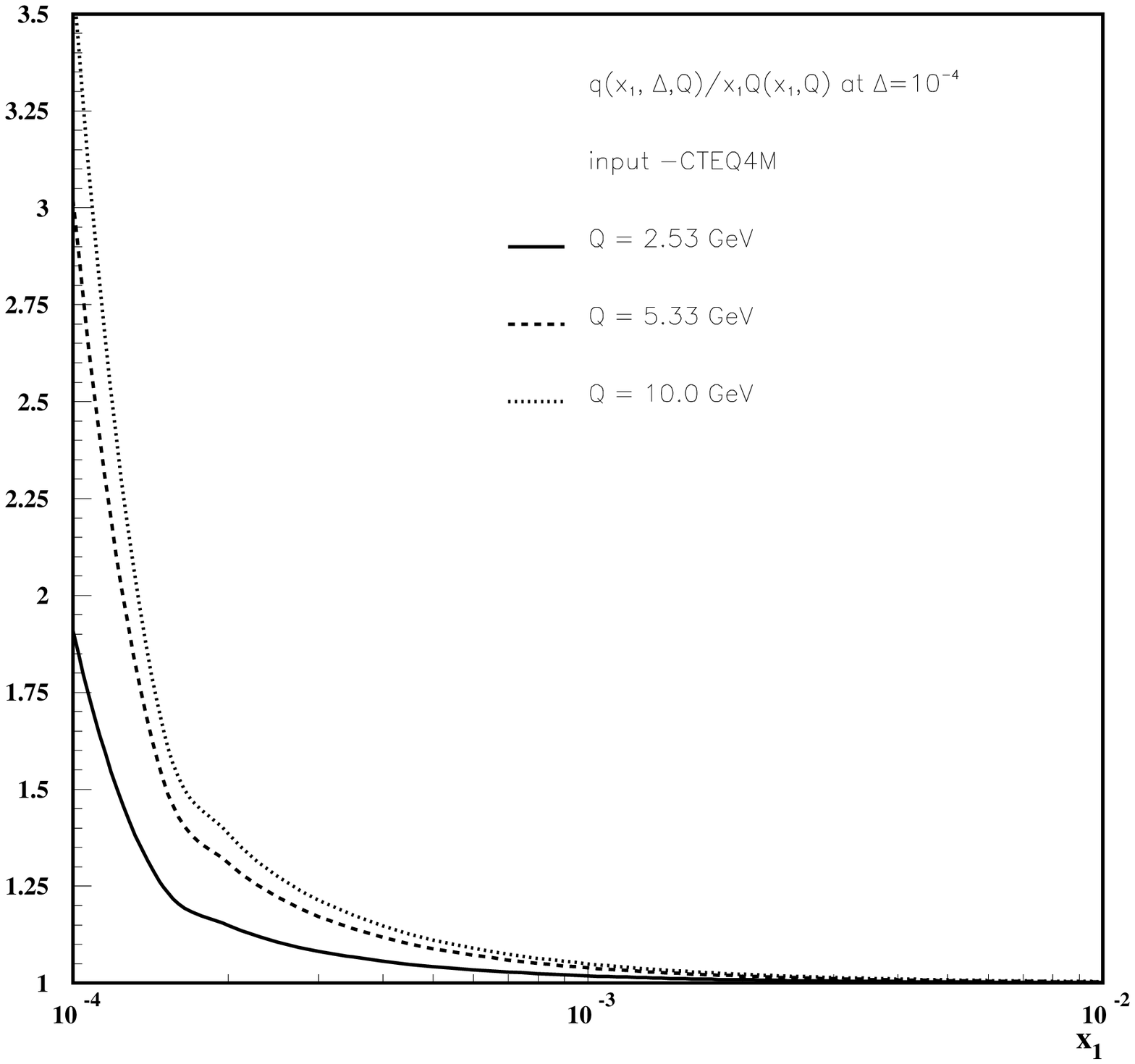,width=14cm,height=15cm}}
\vskip-3cm
\mbox{\epsfig{file=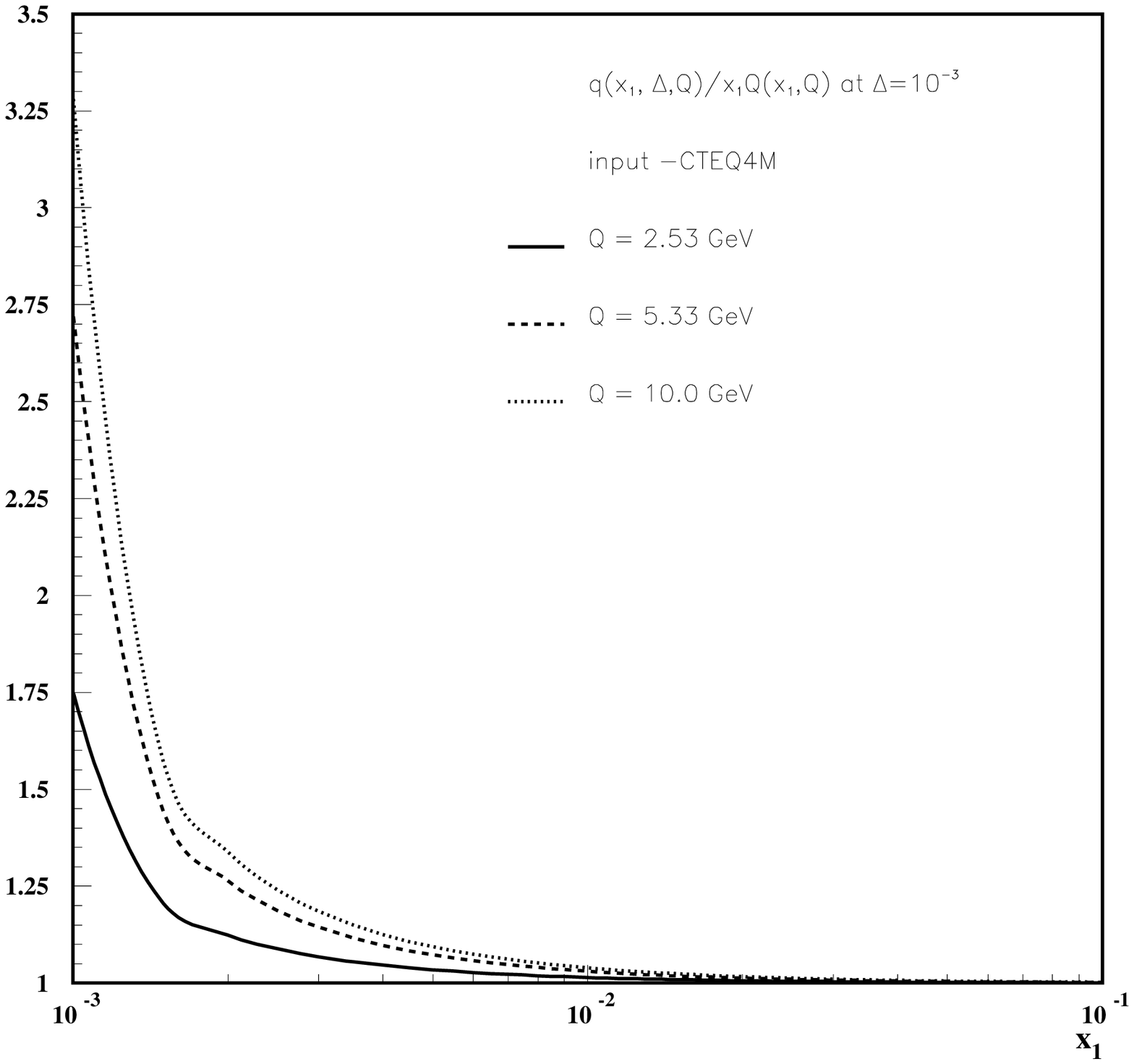,width=14cm,height=15cm}}
\vskip-2cm
\caption{$R_q$ is plotted versus $x_1$ for fixed $\Delta$ using the CTEQ4M 
parameterization with $Q_0=$\mbox{1.6 GeV} and $\Lambda$ =\mbox{202 MeV}.}
\label{nddratio1}
\end{figure}
\newpage
\begin{figure}
\centering
\vskip-3cm
\mbox{\epsfig{file=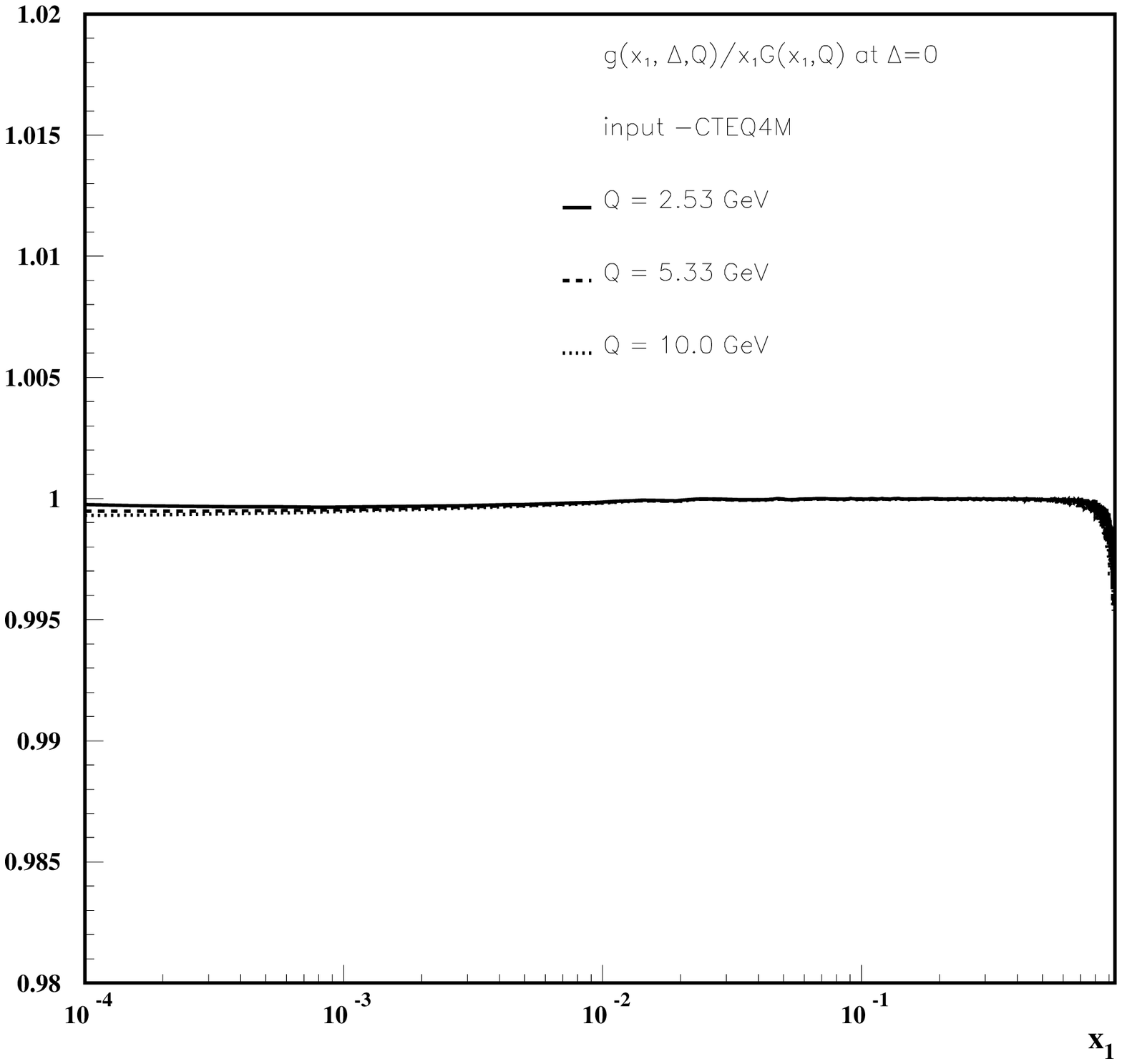,width=14cm,height=15cm}}
\vskip-3cm
\mbox{\epsfig{file=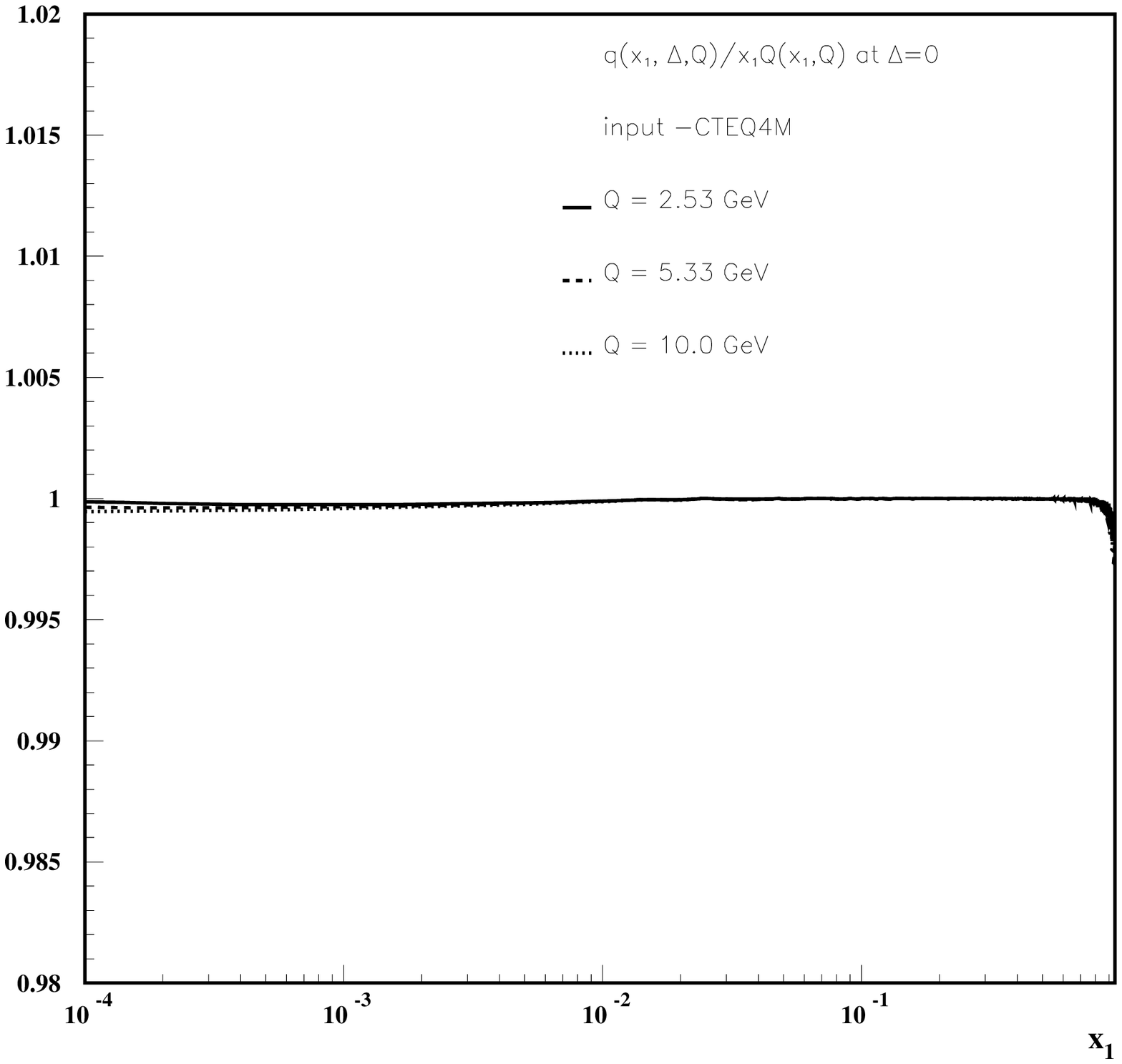,width=14cm,height=15cm}}
\vskip-2cm
\caption{$R_g$ and $R_q$ are plotted versus $x_1$ for $\Delta = 0$ using the 
CTEQ4M parameterization with $Q_0=$\mbox{1.6 GeV} and $\Lambda$ =\mbox{ 202 MeV}.}
\label{nddratio2}
\end{figure}
\newpage
\begin{figure}
\centering
\vskip-3cm
\mbox{\epsfig{file=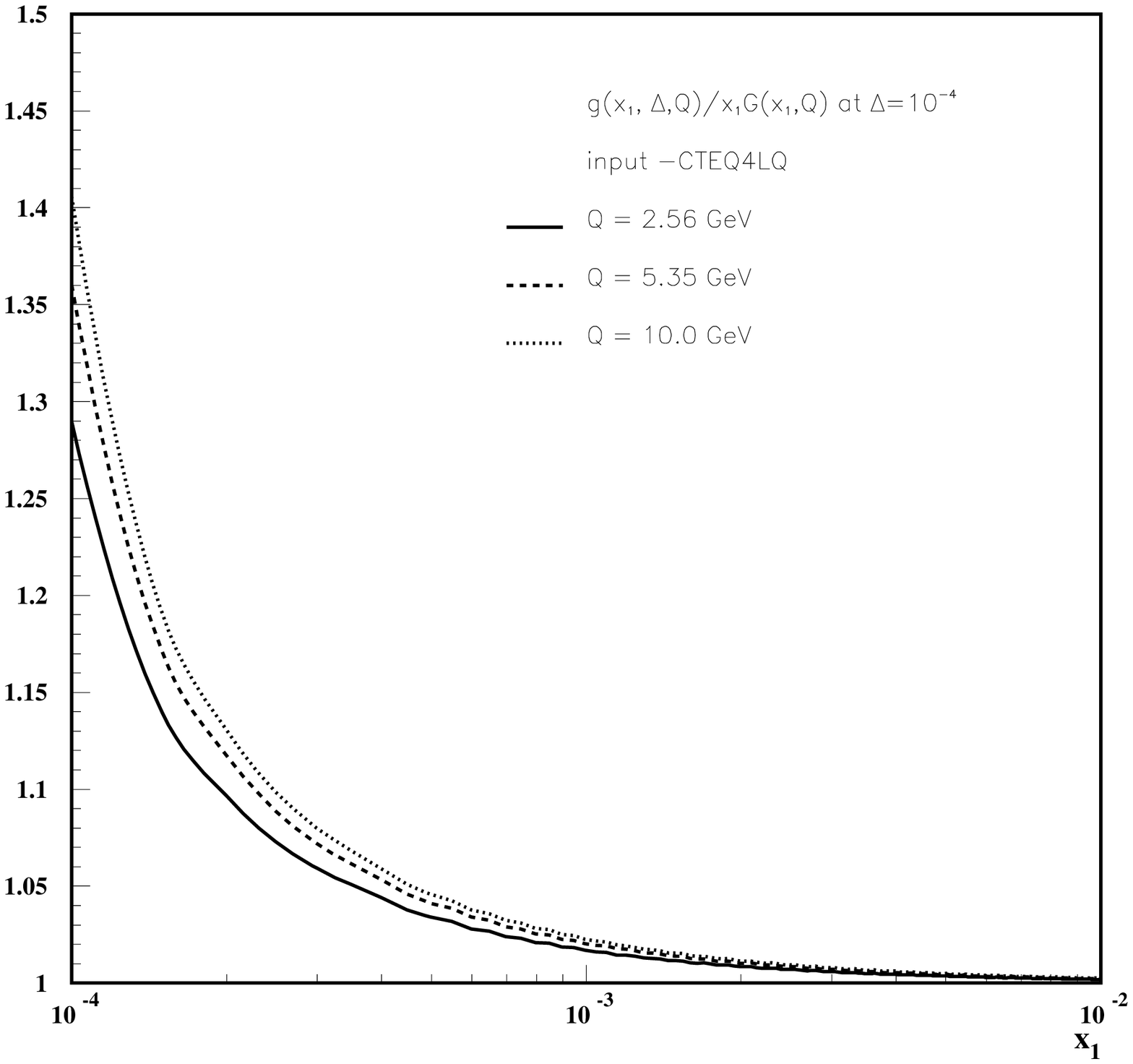,width=14cm,height=15cm}}
\vskip-3cm
\mbox{\epsfig{file=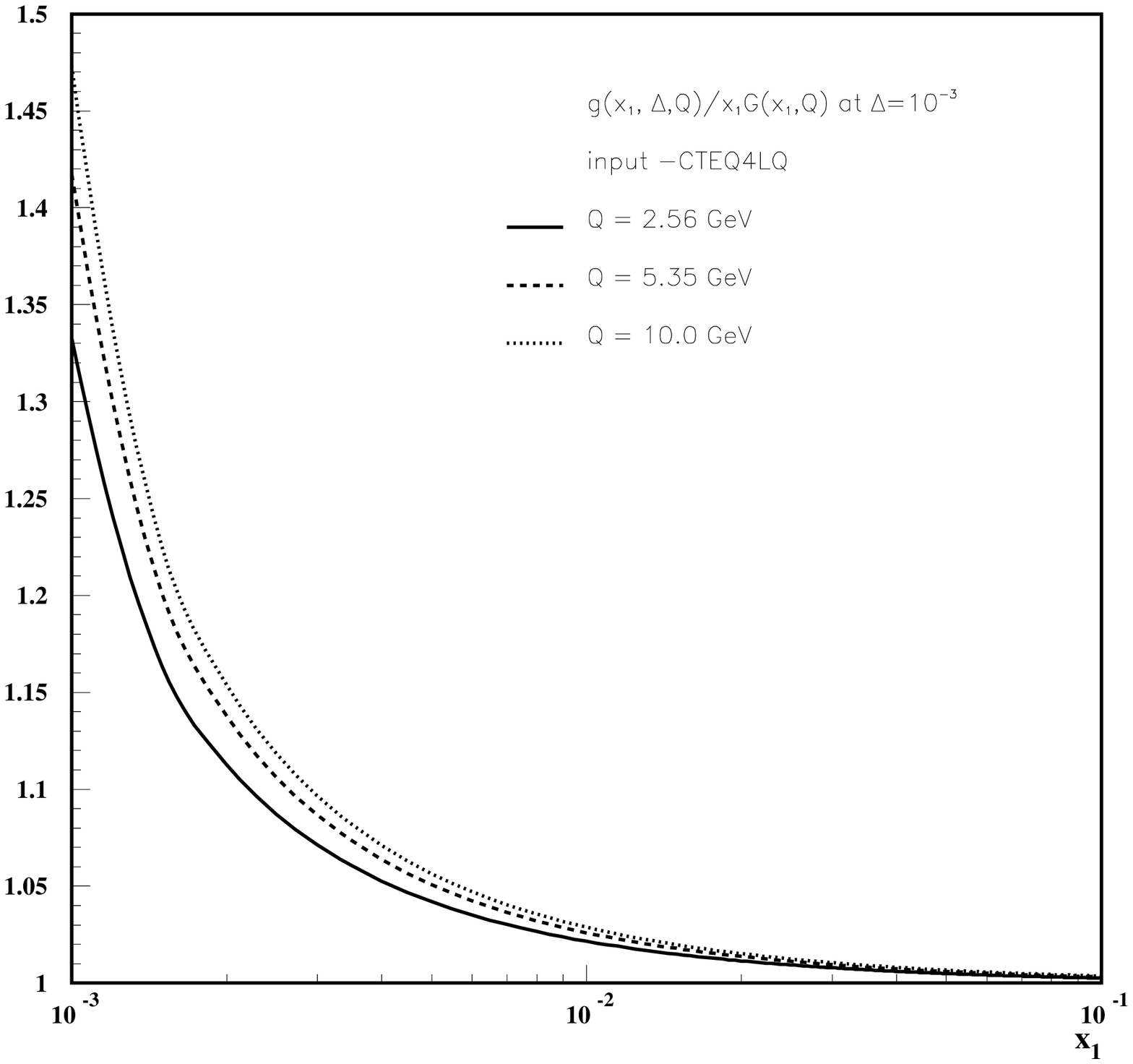,width=14cm,height=15cm}}
\vskip-2cm
\caption{$R_g$ is plotted versus $x_1$ for fixed $\Delta$ using the 
CTEQ4LQ parameterization with $Q_0=$\mbox{0.7 GeV} and 
$\Lambda$ =\mbox{174 MeV}.}
\label{nddratio3}
\end{figure}
\newpage
\begin{figure}
\centering
\vskip-3cm
\mbox{\epsfig{file=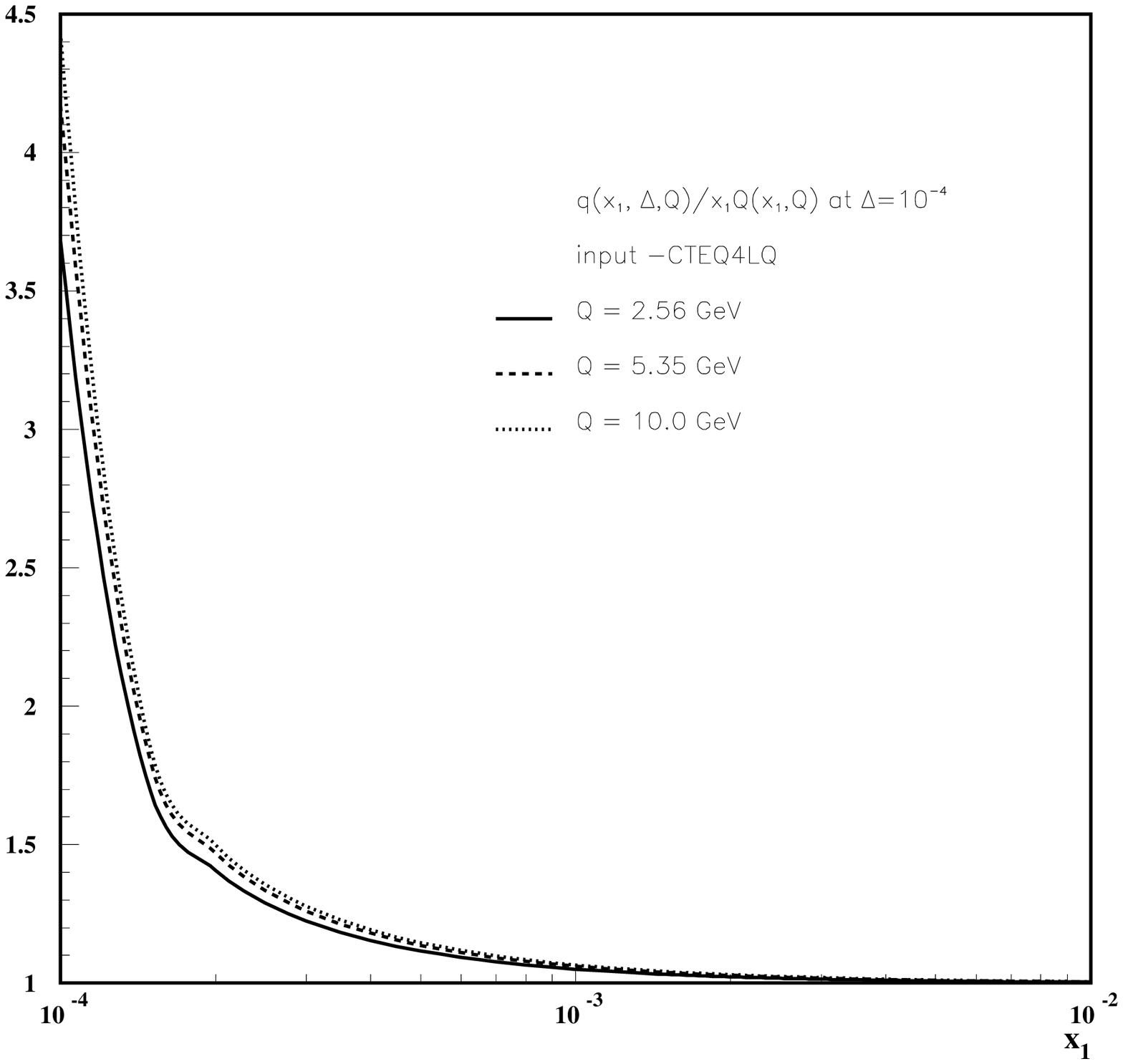,width=14cm,height=15cm}}
\vskip-3cm
\mbox{\epsfig{file=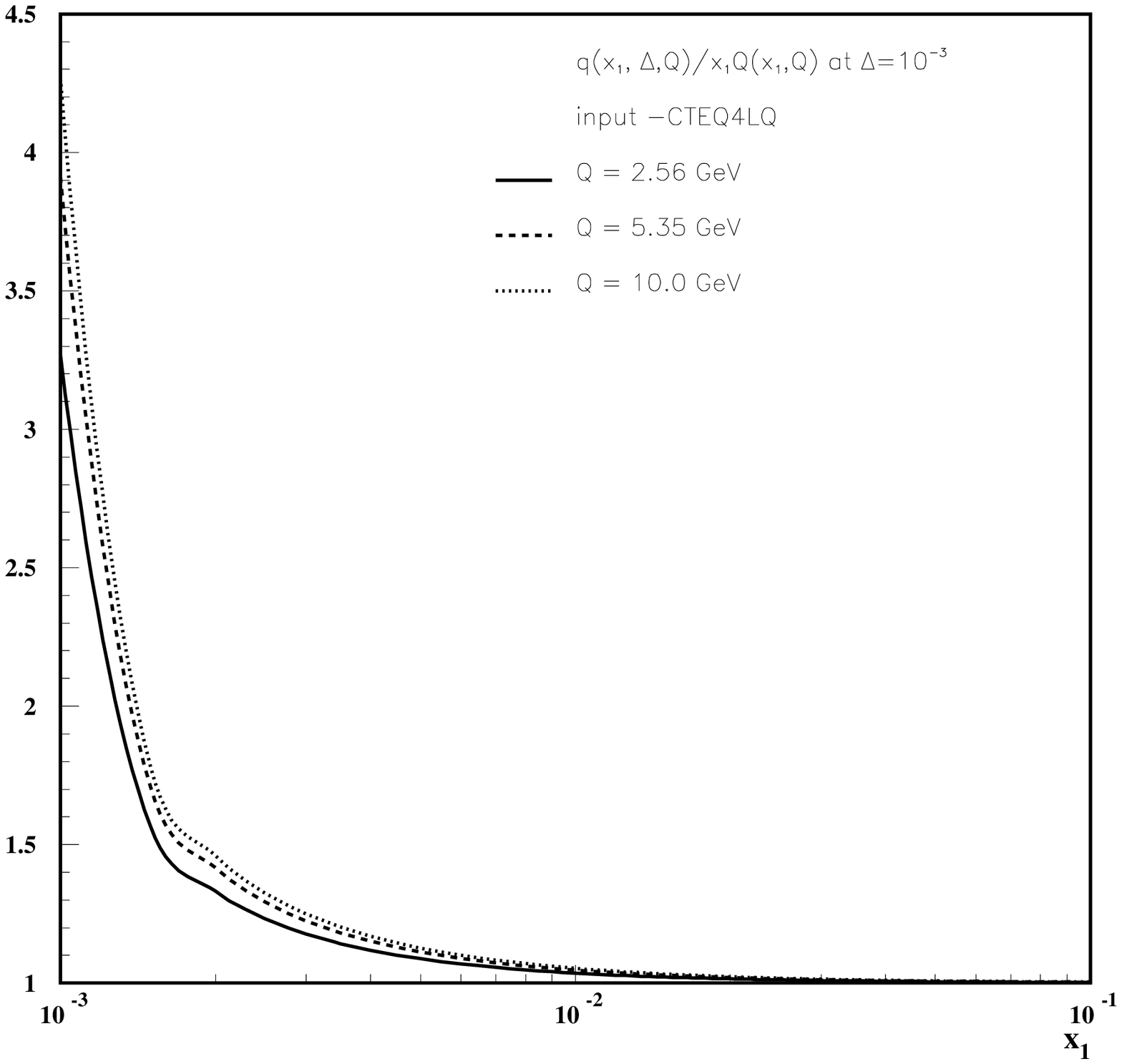,width=14cm,height=15cm}}
\vskip-2cm
\caption{$R_q$ is plotted versus $x_1$ for fixed $\Delta$ using the 
CTEQ4LQ parameterization with $Q_0=$\mbox{0.7 GeV} and 
$\Lambda$ =\mbox{174 MeV}.}
\label{nddratio4}
\end{figure}
\newpage
\begin{figure}
\centering
\vskip-3cm
\mbox{\epsfig{file=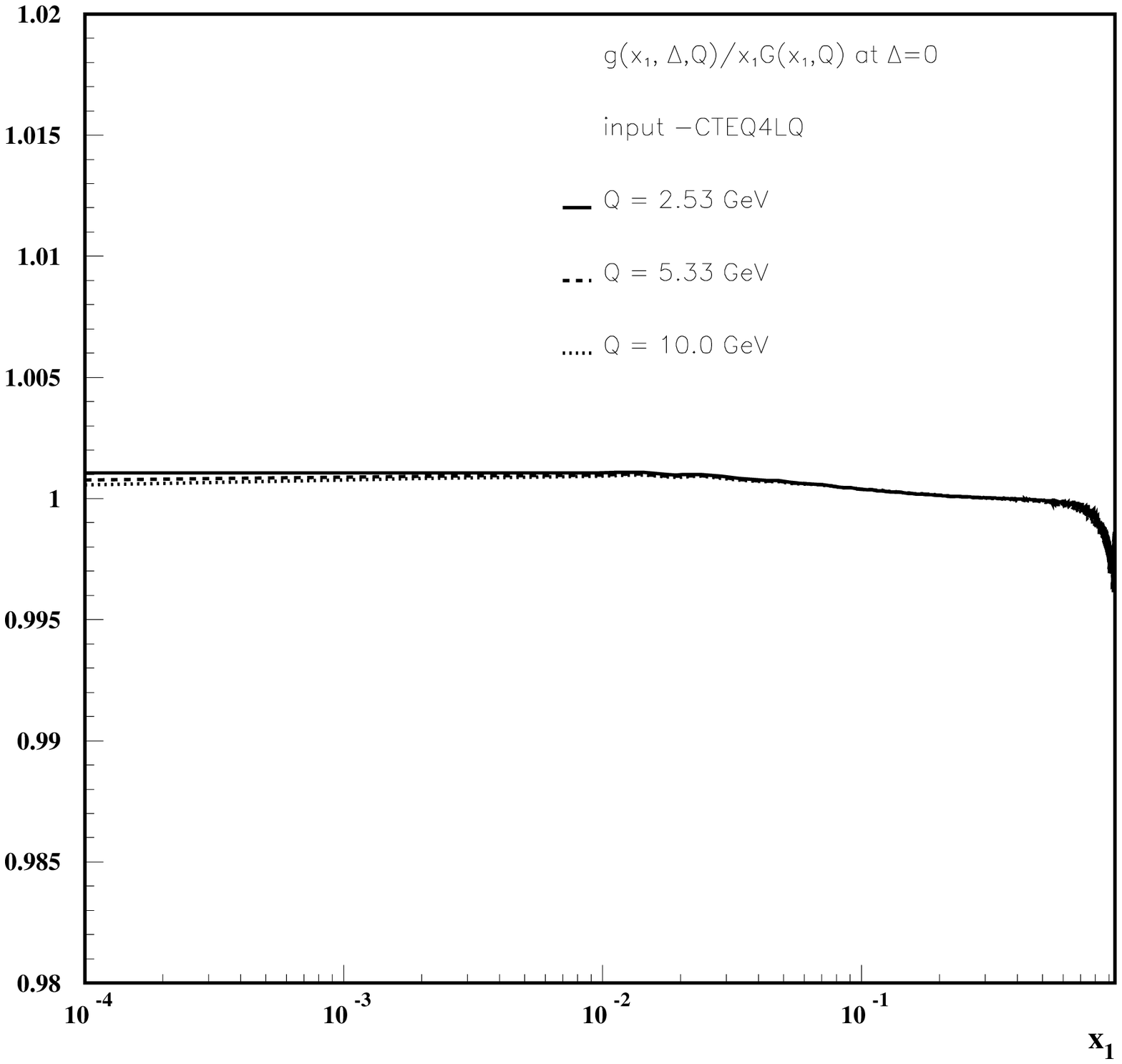,width=14cm,height=15cm}}
\vskip-3cm
\mbox{\epsfig{file=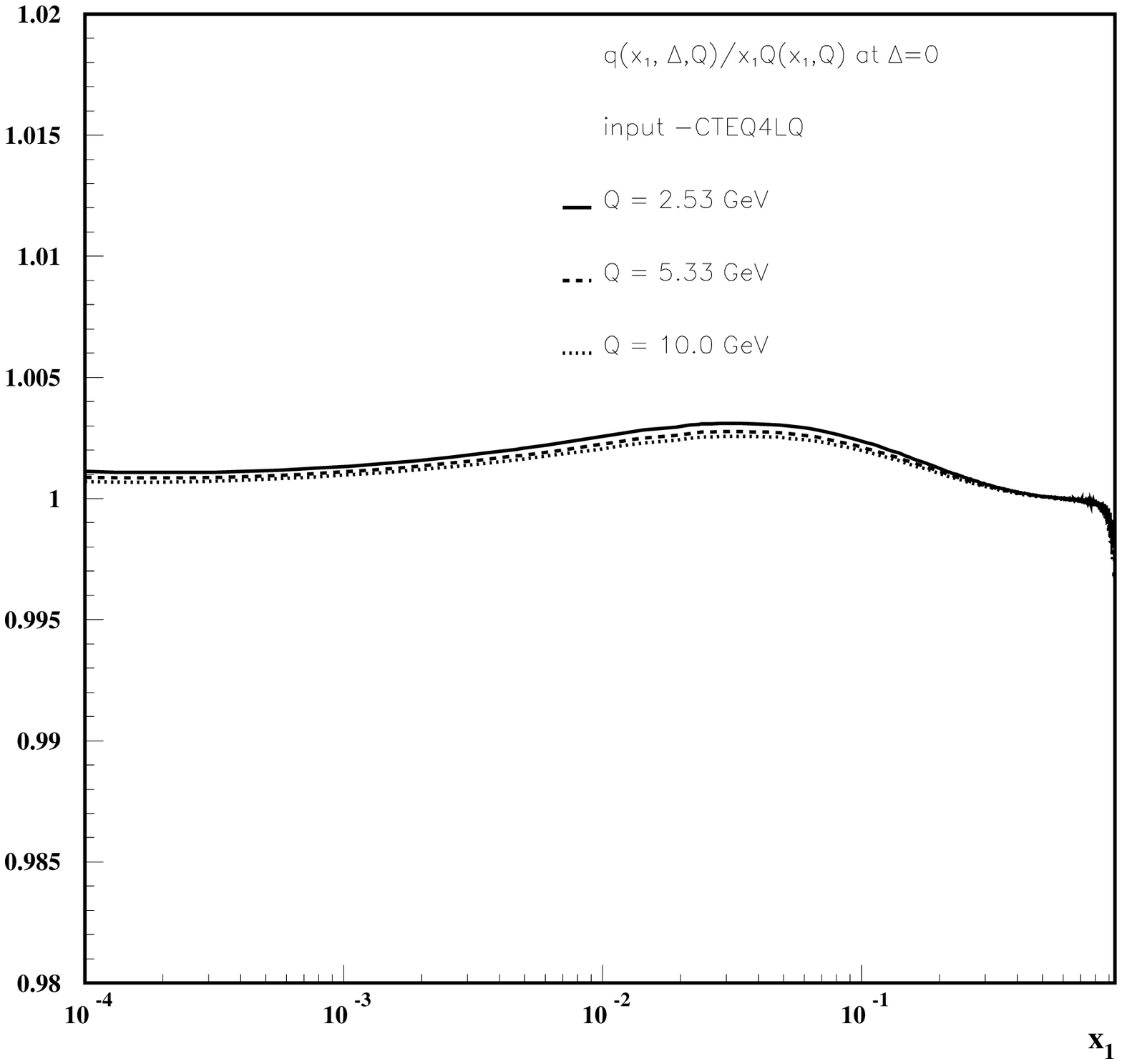,width=14cm,height=15cm}}
\vskip-2cm
\caption{$R_g$ and $R_q$ are plotted versus $x_1$ for $\Delta = 0$ using the 
CTEQ4LQ parameterization with $Q_0=$\mbox{0.7 GeV} and 
$\Lambda$ =\mbox{174 MeV}.}
\label{nddratio5}
\end{figure}
\newpage


\begin{references}
\bibitem {f2} The parton distributions functions are smooth and well behaved 
              thus one just has to use enough points in $x$.
\bibitem {f3} The general analytic expressions for the convolution integrals 
              in an arbitrary $x$-bin were obtained with the help of 
              MATHEMATICA.
\bibitem {f4} The value of $\Delta$ is specified in NINTEGR.
\bibitem {f5} The value of the output at position $N$ is always $0$ since in 
              this case the upper and the lower bound of the integral coincide.
\bibitem {f6} We also plot the same ratio for $\Delta = 0$ to demonstrate the 
              deviation from our code in the diagonal limit from the CTEQ-code.
\bibitem {f8} This was also the case in Ref.\ \cite{ours} where the authors 
              mistakenly put the energies as $Q^2$ where in fact they should 
              have been $Q$, which led to some confusion in the comparisons of
              this first study to Ref.\ \cite{MR97}.
\end{references}
\end{document}